\def\be{\begin{equation}}
\def\ee{\end{equation}}
\def\ba{\begin{array}}
\def\ea{\end{array}}

\def\p{\prime}

\def\Nb{{I\!\! N}}

\def\Cb{\ \hbox{\vrule width 0.6pt height 6pt depth 0pt
              \hskip -3.2 pt} C}
\documentstyle[12pt]{article}
\begin{document}
\parskip=4pt
\parindent=18pt
\baselineskip=22pt \setcounter{page}{1}
\centerline{\Large\bf
Nonlocal Properties and Local Invariants}
\vspace{2ex}
\centerline{\Large\bf for Bipartite Systems}
\vspace{4ex}
\begin{center}
Sergio Albeverio$^{a}$ \footnote{ SFB 611; BiBoS; IZKS; CERFIM(Locarno);
Acc. Arch.; USI(Mendriso)

~~~e-mail: albeverio@uni-bonn.de}, Shao-Ming Fei$^{a,b}$
\footnote{e-mail: fei@uni-bonn.de}, Preeti Parashar$^{c}$
\footnote{e-mail: parashar@isical.ac.in}, Wen-Li Yang$^{d,e}$
\footnote{e-mail: wlyang@yukawa.kyoto-u.ac.jp}

\vspace{3ex}
\begin{minipage}{5.8in}

{\small $~^{a}$ Institut f\"ur Angewandte Mathematik,
Universit\"at Bonn, D-53115}

{\small $~^{b}$ Department of Mathematics, Capital Normal
University, Beijing 100037}

{\small $~^{c}$ Physics and Applied Mathematics Unit, Indian Statistical
Institute, Kolkata 700108}

{\small $~^{d}$ Yukawa Institute for Theoretical Physics, Kyoto
University, Kyoto 606-8502}

{\small $~^{e}$ Institute of Modern Physics, Northwest University,
Xian 710069}

\end{minipage}
\end{center}

\vskip 1 true cm
\parindent=18pt
\parskip=6pt
\begin{center}
\begin{minipage}{5in}
\vspace{3ex} \centerline{\large Abstract} \vspace{4ex}

The nonlocal properties for a kind of generic $N$-dimensional
bipartite quantum systems are investigated. A complete set of
invariants under local unitary transformations is presented. It is
shown that two generic density matrices are locally equivalent if
and only if all these invariants have equal values in these
density matrices.

\bigskip
\medskip
\bigskip
\medskip

\vfill

\end{minipage}
\end{center}

\newpage

As a fundamental phenomenon in quantum mechanics, nonlocality has
been given a lot of attention in foundational considerations,
in the discussion of Bell type inequalities \cite{Bell}
and hidden variable models, see e.g. \cite{ak}. Nonlocality
turns out to be also very important in quantum information
processing such as quantum computation \cite{DiVincenzo}, quantum
teleportation \cite{teleport,teleport1,Ari00,teleport2}, dense
coding \cite{dense} and quantum cryptographic schemes
\cite{crypto1,crypto2,crypto3}. Nonlocal correlations in quantum
systems imply a kind of entanglement among the quantum subsystems.
The nonlocal properties as well as the entanglement of two parts
of a quantum system remain invariant under local transformations of
these parts.

The method developed in \cite{Rains,Grassl}, in principle, allows
one to compute all the invariants of local unitary
transformations, though it is not easy to perform it operationally. In
\cite{makhlin}, the invariants for general two-qubit systems are
studied and a complete set of 18 polynomial invariants is
presented. It is proven that two qubit mixed states are locally
equivalent if and only if all these 18 invariants have equal
values in these states. Therefore any nonlocal characteristics of
entanglement is a function of these invariants. In \cite{linden}
three qubits states are also discussed in detail
from a similar point of view.

In the present paper we discuss the locally invariant properties of
arbitrary dimensional bipartite quantum systems. We present a
complete set of invariants and show that two generic density
matrices with full rank are locally equivalent if and only if all these
invariants have equal values in these density matrices.

We first consider the case of pure states. Let $H$ be an
$N$-dimensional complex Hilbert space, with $\vert i\rangle$, $i=1,...,N$,
as
an orthonormal basis. A general pure state on $H\otimes H$ is of
the form
\begin{equation}\label{mmm}
\vert\Psi\rangle=\sum_{i,j=1}^N a_{ij}\vert i\rangle \otimes
\vert j\rangle,~~~~~~a_{ij}\in\Cb
\end{equation}
with the normalization $\displaystyle\sum_{i,j=1}^N
a_{ij}a_{ij}^\ast=1$ ($\ast$ denoting complex conjugation).

A quantity is called an invariant associated with the
state $\vert\Psi\rangle$ if it is invariant under all local
unitary transformations, i.e. all maps of the form $U\otimes U$
from $H \otimes H$ to
itself, where $U$ is a unitary transformation on the Hilbert space
$H$. Let $A$ denote the matrix given
by $(A)_{ij}=a_{ij}$. The following quantities are known to be
invariants associated with the state $\vert\Psi\rangle$ given by (\ref{mmm}),
see \cite{Linden,afei,qsep,Fan02}:
\begin{equation}\label{I}
I_\alpha=Tr(AA^\dag)^{\alpha},~~~~~~~~~~~\alpha=1,...,N;
\end{equation}
(with $A^\dag$ the adjoint of the matrix $A$).

In terms of the Schmidt decomposition, a given $\vert\Psi\rangle$
can always be written in the following form, using two suitable
orthonormal basis
$\left\{\vert i\rangle^{\prime}\right\}$, $\left\{\vert i\rangle^{\prime
\prime}\right\}$, $i=1,...,N$:
$$
\vert\Psi\rangle=\sum_{i=1}^N \sqrt{\Lambda_i}\vert i\rangle^{\prime}
\otimes
\vert i\rangle^{\prime \prime},
$$
where $\displaystyle\sum_{i=1}^N \Lambda_i=1$, $\Lambda_i\geq 0$.
The $\Lambda_i$, $i=1,...,N$, are the eigenvalues of the matrix $AA^\dag$. As
$AA^\dag$ is self-adjoint, there always exists a unitary matrix
$V$, $VV^\dag=V^\dag V=1$, such that $VAA^\dag
V^\dag=diag\{\Lambda_1,...,\Lambda_N\}$. The invariants (\ref{I}) can then
be written in the form:
$$
I_\alpha=\sum_{i=1}^N \Lambda_i^{\alpha},~~~~~~~\alpha=1,...,N.
$$

As the eigenvalues of the matrix $AA^\dag$ are given by the invariants
under local unitary transformations,
two pure states (on $H\otimes H$) are equivalent under locally unitary
transformations if and only if they have the same values of the
invariants $I_\alpha$, $\alpha=1,...,N$ \cite{Sch06}.
Moreover two Hermitian $m\times m$ matrices $A$ and $B$ are
unitary equivalent ({\it i.e.}, there exists a unitary matrix
$u$ on an $m$-dimensional complex vector space
satisfying $A=uBu^{\dag}$) if and only if
\be\label{Lemma2}
Tr(A^{\alpha})=Tr(B^{\alpha}),~~~{\rm for}~~\alpha=1, ..., m.
\ee

We consider now mixed states on $H\otimes H$. Let $\rho$ be a density
matrix defined on $H\otimes H$ with $rank(\rho)=n\leq N^2$. $\rho$ can
be decomposed according to its eigenvalues and eigenvectors:
$$
\rho=\sum_{i=1}^n\lambda_i\vert\nu_i\rangle\langle\nu_i\vert,
$$
where $\lambda_i$ and $\vert\nu_i\rangle$, $i=1,...,n$, are the
nonzero eigenvalues and eigenvectors respectively of the density matrix $\rho$.
$\vert\nu_i\rangle$ has the form
$$
\vert\nu_i\rangle=\sum_{k,l=1}^N a_{kl}^i \vert k\rangle \otimes
\vert l\rangle,~~~
a_{kl}^i\in\Cb,~~~ \sum_{k,l=1}^N a_{kl}^i a_{kl}^{i\ast}=1,~~~
i=1,...,n.
$$
Let $A_i$ denote the matrix given by $(A_i)_{kl}=a_{kl}^i$. We introduce
$ \left\{\rho_{i}\right\}$,  $ \left\{\theta_{i}\right\}$,
\be\label{rti}
\rho_i=Tr_2 \vert\nu_i\rangle\langle\nu_i\vert=A_iA_i^\dag,~~~
\theta_i=(Tr_1 \vert\nu_i\rangle\langle\nu_i\vert)^\ast=A_i^\dag A_i,
~~~ i,j=1,...,n,
\ee
$Tr_1$ and $Tr_2$ stand for the traces over the first and second
Hilbert spaces respectively, and therefore, $\rho_i$ and $\theta_i$
can be regarded as reduced density matrices.
Let $\Omega(\rho)$ and $\Theta(\rho)$
be two ``metric tensor" matrices, with entries given by
\be\label{ij}
\Omega(\rho)_{ij}=Tr(\rho_i\rho_j),~~~
\Theta(\rho)_{ij}=Tr(\theta_i\theta_j),~~~{\rm for}~ i,j=1,...,n,
\ee
and
$$
\Omega(\rho)_{ij}=\Theta(\rho)_{ij}=0,~~~{\rm for}~ N^2\geq i,j>n.
$$
We call a mixed state $\rho$ a {\it generic} one \footnote{These
states are all the ones but a set of measure zero:
$\{\rho\,|det(\Omega(\rho))=0,~det(\Theta(\rho))=0\}$.} if the
corresponding ``metric tensor" matrices $\Omega$, $\Theta$ satisfy
\be\label{gen} det(\Omega(\rho))\neq 0,~~{\rm
and}~det(\Theta(\rho))\neq 0. \ee Obviously, a generic state
implies $n=N^2$ or $det(\rho)\neq 0$, namely, a state with full
rank. Nevertheless a fully ranked density matrix may be not
generic in the sense of (\ref{gen}).

Similarly we also introduce $X(\rho)$ and $Y(\rho)$ as
\be\label{ijk}
X(\rho)_{ijk}=Tr(\rho_i\rho_j\rho_k),~~~
Y(\rho)_{ijk}=Tr(\theta_i\theta_j\theta_k),~~~~ i,j,k=1,...,n.
\ee

{\sf[Theorem].} Two generic density matrices with full rank are equivalent
under
local unitary transformations if and only if there exists a ordering of
the corresponding eigenstates such that the following
invariants have the same values for both density matrices:
\be\label{theorem}
\ba{l}
J^s(\rho)=Tr_2(Tr_1\rho^s),~~~s=1,...,N^2;\\[4mm]
\Omega(\rho),~~~\Theta(\rho),~~~
X(\rho),~~~ Y(\rho).
\ea
\ee

\noindent {\bf Remark 1.} It is well-known that the set of
eigenvalues and corresponding eigenstates is uniquely defined, but
not the labelling of them. However, from the proof below, one can
see that two generic density matrices would have the same set of
eigenvalues if they share the same values $\{ J^{s}(\rho)\}$. One
can uniquely choose the label for the eigenstates with the
different eigenvalues. For the case of degenerate eigenvalues,
if two generic density matrices $\rho$ and $\rho^\prime$ are
equivalent under the local
unitary transformations, one can always find a kind of
label for the eigenstates such that they share
the same invariants (\ref{theorem}), i.e., under this label,
$\Omega(\rho)_{ij}=\Omega(\rho^\prime)_{ij}$,
$\Theta(\rho)_{ij}=\Theta(\rho^\prime)_{ij}$,
$X(\rho)_{ijk}=X(\rho^\prime)_{ijk}$,
$Y(\rho)_{ijk}=Y(\rho^\prime)_{ijk}$. This is due to that
these invariants are the sufficient and necessary conditions
for two generic density matrices to be equivalent under local
unitary transformations, see the proof below.

{\sf[Proof].} We first show that the quantities given in
(\ref{theorem}) are invariant under local unitary transformations. Let
$u$ and $w$ be unitary transformations,
$uu^\dag=u^\dag u=ww^\dag=w^\dag w=1$. Under the local
unitary transformation $u\otimes w$, we have
$\rho\to\rho^\prime=u\otimes w ~\rho~u^\dag\otimes w^\dag$.
Correspondingly, we have $\vert\nu_i\rangle
\to \vert\nu_i^\prime\rangle=u\otimes w\vert\nu_i\rangle$, or
equivalently $A_i$ is mapped to $A_i^\prime=u^t A_i w$, where
$u^t$ is the transpose of $u$. Therefore
\be\label{aa}
\rho_i^\prime=A_i^\prime
A_i^{\prime\dag}=u^t A_i A_i^\dag u^\ast=u^t \rho_i u^\ast,~~~
\theta_i^\prime=A_i^{\prime\dag}A_i^\prime=w^\dag A_i^\dag A_i w
=w^\dag \theta_i w.
\ee
By using (\ref{aa}), it is straightforward to check the following relations:
$$
\ba{l}\displaystyle
J^s(\rho) \to J^s(\rho^\prime)=Tr_2[\sum_{i=1}^n\lambda_i^s
Tr_1(\vert\nu_i^\prime\rangle\langle\nu_i^\prime\vert)]
=Tr_2[\sum_{i=1}^n\lambda_i^s A_i^\prime
A_i^{\prime\dag}]=J^s(\rho),\\[4mm]
\Omega(\rho)_{ij}\to \Omega(\rho^\prime)_{ij}
=Tr(\rho_i^\prime\rho_j^\prime)
=Tr(u^t \rho_i\rho_j u^\ast)=\Omega(\rho)_{ij},\\[4mm]
\Theta(\rho)_{ij}\to \Theta(\rho^\prime)_{ij}=
Tr(\theta_i^\prime\theta_j^\prime)
=Tr(w^\dag \theta_i\theta_j w)=\Theta(\rho)_{ij}\\[4mm]
X(\rho)_{ijk}\to X(\rho^\prime)_{ijk}
=Tr(\rho_i^\prime\rho_j^\prime\rho_k^\prime)
=Tr(u^t \rho_i\rho_j\rho_k u^\ast)=X(\rho)_{ijk},\\[4mm]
Y(\rho)_{ijk}\to Y(\rho^\prime)_{ijk}=
Tr(\theta_i^\prime\theta_j^\prime\theta_k^\prime)
=Tr(w^\dag \theta_i\theta_j\theta_k w)=Y(\rho)_{ijk},
\ea
$$
where $i,j,k=1,...,n$. Hence the quantities in
(\ref{theorem}) are invariants of local unitary transformations.
If two density matrices are equivalent under local unitary
transformations, then their corresponding invariants in
(\ref{theorem}) have the same values.

Now suppose conversely that the states $\rho=\sum_{i=1}^n\lambda_i\vert
\nu_i\rangle\langle\nu_i\vert$ and $\rho^\prime=
\sum_{i=1}^n\lambda_i^\prime\vert\nu_i^\prime\rangle
\langle\nu_i^\prime\vert$ give the
same values to the invariants in (\ref{theorem}). We are going to
prove that $\rho$ and $\rho^\prime$ are equivalent under
local unitary transformations.

a) As
$$
J^s(\rho)=Tr_2(\sum_{i=1}^n\lambda_i^s
Tr_1(\vert\nu_i\rangle\langle\nu_i\vert))
=Tr_2(\sum_{i=1}^n\lambda_i^s A_iA_i^{\dag})=\sum_{i=1}^n\lambda_i^s,
$$
from $J^s(\rho^\prime)=J^s(\rho)$ we have
$$
\sum_{i=1}^n\lambda_i^{\prime s}=\sum_{i=1}^n\lambda_i^s,~~~~~\forall~
s=1,...,N^2.
$$
From (\ref{Lemma2}),  we have that  $\rho^\prime$ and $\rho$ have
the same nonzero eigenvalues, i.e.,
$\lambda^\prime_i=\lambda_i$, $i=1,...,n$.

b) From (\ref{ij}), the generic condition $det(\Omega(\rho))\neq
0$ implies that $\{\rho_i\}$, $i=1,...,n(=N^2)$, span the space of
$N\times N$
matrices and
\be\label{a}
\rho_i\rho_j=\sum_{k=1}^n C_{ij}^k\rho_k,~~~~~
C_{ij}^k~\in\Cb
\ee
Taking trace of (\ref{a}) and using the condition $Tr\rho_i=1$
one gets
\be\label{b}
\Omega_{ij}=\sum_{k=1}^n C_{ij}^k.
\ee
From (\ref{b}) and (\ref{ijk}) we obtain
$$
X_{ijk}=\sum_{l=1}^n C_{ij}^l\Omega_{lk}.
$$
Therefore
\be\label{d} C_{ij}^l=\sum_{k=1}^n
X_{ijk}\Omega^{lk},
\ee
where the matrices $\Omega^{ij}$ is the corresponding inverses of the
matrices $\Omega_{ij}$ (which exist due to the assumption (\ref{gen})).
(\ref{d}) implies that the  coefficients $C_{ij}^l$ are given by
$\{\Omega_{ij}, X_{ijk}\}$.
From (\ref{ij}), the generic condition (\ref{gen}) implies
that $\{\rho_i\}$ forms an irreducible $N$-dimensional representation of
the algebra $gl(N,\Cb)$ with the generators $\{e_i, ~~i=1,..., N^2\}$
satisfying
\be\label{Alg}
[e_i, e_j]=\sum_{k=1}^{N^2}f_{ij}^k e_k,
\ee
where $f_{ij}^k=C_{ij}^k-C_{ji}^k$. More explicitly,
$\pi\left(e_i\right)=\rho_i$, $i=1,...,N^2$, where $\pi$ is the
representation of $gl(N,\Cb)$.

The generic condition $det(\Omega(\rho^{\prime}))\neq 0$ implies
that $\{\rho^{\prime}_i\}$, $i=1,...,N^2$, also span the space of
$N\times N$ matrices,
\be\label{e}
\rho_i^\p\rho_j^\p=\sum_{k=1}^n
C_{ij}^{\p k} \rho_k^\p,~~~~~C_{ij}^{\p k}\in\Cb.
\ee
If $\Omega(\rho^\p)=\Omega(\rho)$ and $X(\rho^\p)=X(\rho)$,  we have
$C_{ij}^{\p l}=C_{ij}^l$. Therefore, $\left\{\rho_i\right\}$ and
$\left\{\rho^{\prime}_i\right\}$ (if one chooses
$\pi^{\prime}(e_i)=\rho^{\prime}_i$) are two irreducible
$N$-dimensional representation of $gl(N, \Cb)$ (\ref{Alg}). It is
well-known that all the Casimir operators of the algebra
$gl(N,\Cb)$ can be expressed in terms of homogeneous polynomials of
$e_i$'s (for example, the first Casimir operator $C_2$ can be
written as a quadratic polynomial of $e_i$'s).
Moreover Casimir operators
are algebraically independent and give rise to a complete set of
generators for the center of the universal enveloping algebra of
$gl(N, \Cb)$. They take scalar values on an irreducible
representation of $gl(N,\Cb)$ (from Schur's Lemma), and become the
characters of the irreducible representations \cite{Hum72}.
Due to the fact that the
trace of every polynomial of $\{\rho_i \}$ and
$\{\rho^{\prime}_i\}$ can be represented in terms of
$\{\Omega_{ij}(\rho),~X_{ijk}(\rho)\}$ and
$\{\Omega_{ij}(\rho^{\prime}),~Y_{ijk}(\rho^{\prime})\}$ respectively (see
below remark 2), we conclude that the values of all the
Casimir operators given by the two $N$-dimensional representations
$\left\{\rho_i\right\}$ and $\left\{\rho^{\prime}_i\right\}$ are
equal, from the condition $\Omega(\rho)=\Omega(\rho^{\prime})$ and
$X_{ijk}(\rho)=X_{ijk}(\rho^{\prime})$. Hence, the two sets of
representations (primed and unprimed) of the algebra $gl(N,\Cb)$
are equivalent, i.e., \be\label{f} \rho_i^\p=u^t\rho_i u^\ast, \ee
for some $u\in{\cal U}$.

Similarly, from $\Theta(\rho)=\Theta(\rho^{\prime})$ and
$Y_{ijk}(\rho)=Y_{ijk}(\rho^{\prime})$ we can deduce that
\be\label{f1}
\theta_i^\p=w^{\dag}\theta_i w,~~{\rm for~some~} w\in{\cal U}.
\ee
From the {\it Singular value decomposition} of matrices \cite{Nie00}, we
have $\vert\nu_i^\prime\rangle=u\otimes w \vert\nu_i\rangle$,
$i=1,...,N^2$,
and $\rho^\prime=u\otimes w ~\rho~u^\dag\otimes w^\dag$. Hence $\rho^\prime$
and $\rho$ are equivalent under local unitary transformations.
\hfill $\rule{2mm}{2mm}$

\noindent{\bf Remark 2}. For a degenerate state $\rho$,
$det(\Omega(\rho))=0 $ (resp. $det(\Theta(\rho))=0$), the above
invariants (\ref{theorem}) are not complete in the sense that two
degenerate density matrices can be not equivalent under local
unitary transformations even if they give the same values to the
invariants in (\ref{theorem}). This is due to the fact
that there exist null vectors for the degenerate state.
For example in the case $det(\Omega(\rho))=0$, there exists at
least one Hermitian matrix $B$ which satisfies $Tr(B\rho_i)=0$ for
$i=1, ..., n$. Hence $\Omega(\rho^\prime)_{ij}=\Omega(\rho)_{ij}$
and $X(\rho^\p)_{ijk}=X(\rho)_{ijk}$
are not enough to get the first equivalence relation (\ref{f}).
In this case
some new invariants have to be introduced to get a complete set
of invariants.
From the algebraic relations (\ref{a}) and formula (\ref{d}),
other generalized invariants
like $Tr((\rho_i)^{m_i}(\rho_j)^{m_j}...(\rho_k)^{m_k})$ and
$Tr((\theta_i)^{m_i}(\theta_j)^{m_j}...(\theta_k)^{m_k})$,
$i,j,...,k=1,...,n$; $m_i,m_j,...,m_k\in\Nb$,
can be represented in terms of
$\{\Omega_{ij},
X_{ijk}\}$ and  $\{\Theta_{ij}, Y_{ijk}\}$ for a generic state with full
rank,  for example,
$$
Tr\left(\rho_{i_1}\rho_{i_2} \cdots \rho_{i_m}\right)
=\sum_{\{\alpha_1,\cdots,\alpha_{m-2}\}}C^{\alpha_1}_{i_1~i_2}
C^{\alpha_2}_{\alpha_1~i_3} \cdots
C^{\alpha_{m-2}}_{\alpha_{m-3}~i_{m-1}}\Omega_{\alpha_{m-2}~i_m}.
$$
Hence by doing so we do not get new invariants.

To summarize, we have discussed here the local invariants for
arbitrary dimensional bipartite quantum systems and have presented
a set of invariants of local unitary transformations. The set of
invariants is not necessarily independent (they could be
represented by each other in some cases) but it is complete in the
sense that two generic density matrices are equivalent under local
unitary transformations if and only if all these invariants have
equal values for these density matrices.

\vspace{1.0truecm}

\noindent {\bf Acknowledgments} We would like to thank M. Grassl
for interesting discussions. P.P. is grateful to Prof. Yu. I.
Manin for encouragement and the Max-Planck Institute f\"ur
Mathematik, Bonn, for warm hospitality, where a part of this work
was done. W-L Yang would like to thank Prof. R. Sasaki and acknowledge the
Yukawa Institute for Theoretical Physics, Kyoto University and the JSPS for
their warm hospitality.

\end{document}